\documentstyle[12pt,a4]{article}

\begin{document}
\renewcommand{\theequation}{\thesection.\arabic{equation}}
\thispagestyle{empty}
\vspace*{-1.5cm}
\hfill {\small KL--TH 97/4} \\[8mm]

\message{reelletc.tex (Version 1.0): Befehle zur Darstellung |R  |N, Aufruf z.B. \string\bbbr}
%
%
\message{reelletc.tex (Version 1.0): Befehle zur Darstellung |R  |N, Aufruf z.B. \string\bbbr}
%
%
%
%
%
\font \smallescriptscriptfont = cmr5
\font \smallescriptfont       = cmr5 at 7pt
\font \smalletextfont         = cmr5 at 10pt
\font \tensans                = cmss10
\font \fivesans               = cmss10 at 5pt
\font \sixsans                = cmss10 at 6pt
\font \sevensans              = cmss10 at 7pt
\font \ninesans               = cmss10 at 9pt
\newfam\sansfam
\textfont\sansfam=\tensans\scriptfont\sansfam=\sevensans
\scriptscriptfont\sansfam=\fivesans
\def\sans{\fam\sansfam\tensans}
\def\bbbr{{\rm I\!R}} 
\def\bbbn{{\rm I\!N}} 
\def\bbbE{{\rm I\!E}} 
\def\bbbm{{\rm I\!M}}
\def\bbbh{{\rm I\!H}}
\def\bbbk{{\rm I\!K}}
\def\bbbd{{\rm I\!D}}
\def\bbbp{{\rm I\!P}}
\def\bbbone{{\mathchoice {\rm 1\mskip-4mu l} {\rm 1\mskip-4mu l}
{\rm 1\mskip-4.5mu l} {\rm 1\mskip-5mu l}}}
\def\bbbc{{\mathchoice {\setbox0=\hbox{$\displaystyle\rm C$}\hbox{\hbox
to0pt{\kern0.4\wd0\vrule height0.9\ht0\hss}\box0}}
{\setbox0=\hbox{$\textstyle\rm C$}\hbox{\hbox
to0pt{\kern0.4\wd0\vrule height0.9\ht0\hss}\box0}}
{\setbox0=\hbox{$\scriptstyle\rm C$}\hbox{\hbox
to0pt{\kern0.4\wd0\vrule height0.9\ht0\hss}\box0}}
{\setbox0=\hbox{$\scriptscriptstyle\rm C$}\hbox{\hbox
to0pt{\kern0.4\wd0\vrule height0.9\ht0\hss}\box0}}}}

\def\bbbe{{\mathchoice {\setbox0=\hbox{\smalletextfont e}\hbox{\raise
0.1\ht0\hbox to0pt{\kern0.4\wd0\vrule width0.3pt height0.7\ht0\hss}\box0}}
{\setbox0=\hbox{\smalletextfont e}\hbox{\raise
0.1\ht0\hbox to0pt{\kern0.4\wd0\vrule width0.3pt height0.7\ht0\hss}\box0}}
{\setbox0=\hbox{\smallescriptfont e}\hbox{\raise
0.1\ht0\hbox to0pt{\kern0.5\wd0\vrule width0.2pt height0.7\ht0\hss}\box0}}
{\setbox0=\hbox{\smallescriptscriptfont e}\hbox{\raise
0.1\ht0\hbox to0pt{\kern0.4\wd0\vrule width0.2pt height0.7\ht0\hss}\box0}}}}

\def\bbbq{{\mathchoice {\setbox0=\hbox{$\displaystyle\rm Q$}\hbox{\raise
0.15\ht0\hbox to0pt{\kern0.4\wd0\vrule height0.8\ht0\hss}\box0}}
{\setbox0=\hbox{$\textstyle\rm Q$}\hbox{\raise
0.15\ht0\hbox to0pt{\kern0.4\wd0\vrule height0.8\ht0\hss}\box0}}
{\setbox0=\hbox{$\scriptstyle\rm Q$}\hbox{\raise
0.15\ht0\hbox to0pt{\kern0.4\wd0\vrule height0.7\ht0\hss}\box0}}
{\setbox0=\hbox{$\scriptscriptstyle\rm Q$}\hbox{\raise
0.15\ht0\hbox to0pt{\kern0.4\wd0\vrule height0.7\ht0\hss}\box0}}}}

\def\bbbt{{\mathchoice {\setbox0=\hbox{$\displaystyle\rm
T$}\hbox{\hbox to0pt{\kern0.3\wd0\vrule height0.9\ht0\hss}\box0}}
{\setbox0=\hbox{$\textstyle\rm T$}\hbox{\hbox
to0pt{\kern0.3\wd0\vrule height0.9\ht0\hss}\box0}}
{\setbox0=\hbox{$\scriptstyle\rm T$}\hbox{\hbox
to0pt{\kern0.3\wd0\vrule height0.9\ht0\hss}\box0}}
{\setbox0=\hbox{$\scriptscriptstyle\rm T$}\hbox{\hbox
to0pt{\kern0.3\wd0\vrule height0.9\ht0\hss}\box0}}}}

\def\bbbs{{\mathchoice
{\setbox0=\hbox{$\displaystyle     \rm S$}\hbox{\raise0.5\ht0\hbox
to0pt{\kern0.35\wd0\vrule height0.45\ht0\hss}\hbox
to0pt{\kern0.55\wd0\vrule height0.5\ht0\hss}\box0}}
{\setbox0=\hbox{$\textstyle        \rm S$}\hbox{\raise0.5\ht0\hbox
to0pt{\kern0.35\wd0\vrule height0.45\ht0\hss}\hbox
to0pt{\kern0.55\wd0\vrule height0.5\ht0\hss}\box0}}
{\setbox0=\hbox{$\scriptstyle      \rm S$}\hbox{\raise0.5\ht0\hbox
to0pt{\kern0.35\wd0\vrule height0.45\ht0\hss}\raise0.05\ht0\hbox
to0pt{\kern0.5\wd0\vrule height0.45\ht0\hss}\box0}}
{\setbox0=\hbox{$\scriptscriptstyle\rm S$}\hbox{\raise0.5\ht0\hbox
to0pt{\kern0.4\wd0\vrule height0.45\ht0\hss}\raise0.05\ht0\hbox
to0pt{\kern0.55\wd0\vrule height0.45\ht0\hss}\box0}}}}

\def\bbbz{{\mathchoice {\hbox{$\sans\textstyle Z\kern-0.4em Z$}}
{\hbox{$\sans\textstyle Z\kern-0.4em Z$}}
{\hbox{$\sans\scriptstyle Z\kern-0.3em Z$}}
{\hbox{$\sans\scriptscriptstyle Z\kern-0.2em Z$}}}}
\setlength{\topmargin}{-1.5cm}
\begin{center}
{\large\bf THE CONTINUOUS SERIES OF CRITICAL POINTS \\
OF THE TWO-MATRIX MODEL AT \\
${\bf N \to \infty}$ \\
IN THE DOUBLE SCALING LIMIT}\\
\vspace{0.5cm}
{\large S. Balaska, J. Maeder and W. R\"uhl}\\
{\it Department of Physics, University of Kaiserslautern, P.O.Box 3049\\
67653 Kaiserslautern, Germany \\
E-mail: ruehl@physik.uni-kl.de}\\
\vspace{5cm}
\begin{abstract}
The critical points of the continuous series are characterized by two complex
numbers $l_{1,2} (Re(l_{1,2})< 0)$, and a natural number $n(n \ge 3)$ which 
enters the string susceptibility constant through $\gamma = \frac{-2}{n-1}$.
The critical potentials are analytic functions with a convergence radius
depending on $l_1$ (or $l_2$). We use the orthogonal
polynomial method and solve the Schwinger-Dyson equations with a technique 
borrowed from conformal field theory.
\end{abstract}
\vspace{3cm}
{\it March 1997}
\end{center}
\newpage
\section{Introduction}
We study matrix models whose action depends on hermitean $N \times N$ matrices
as dynamical variables. They are coupled to a chain of $r$ vertices and
$r-1$ connecting links
\begin{eqnarray}
S(M^{(1)}, M^{(2)}, M^{(3)} \ldots M^{(r)}) = \nonumber \\
= Tr \left\{ \sum^r_{\alpha =1} V_{\alpha}(M^{(\alpha)}) - 
\sum^{r-1}_{\alpha =1} c_{\alpha} M^{(\alpha)} M^{(\alpha+1)} \right\}
\label{1.1}
\end{eqnarray}
Little is known about models with $r \ge 3$, but the two-matrix models
$(r=2)$ seem to exhibit the full richness of critical structures. It turns out 
that they possess two series of critical points: the well-known ``discrete series''
for which the potentials $V_{\alpha}$ are polynomials, and a ``continuous
series'' which we will describe in this work.

Statistical ensembles of matrices appeared first in connection with problems of nuclear
physics \cite{1}. As generalizations of vector sigma models they served as 
objects for the study of phase transitions and renormalization theory
\cite{2}. Recently they attracted interest as models for the coupling of
conformal field matter with the gravitational field \cite{3}. In this case
they are analyzed in their critical domain defined by the ``double scaling
limit''. We shall also apply this limiting procedure in this work.

All investigations of the matrix models in the double scaling limit are
based on the orthogonal polynomial method which will be outlined at the 
end of this introduction. If the critical potentials are polynomials, the
matrix models can be solved perturbatively in the double scaling domain.
This method has been applied to study all types of critical behaviour of
the polynomial two-matrix models in \cite{3,4}. The final result can be
described as follows. Let the polynomial degrees of the potentials be $l_1$ and
$l_2, \, l_1 \le l_2$. If $l_2$ does not divide $l_1$, the universality 
class of the maximal critical point of this model is
\begin{equation}
[p,q] = [l_1,l_2]
\label{1.2}
\end{equation}
where $p$ and $q$ denote the degrees of differential operators of the
generalized KdV hierarchy. If,
however, $l_2$ divides $l_1$, but differs from two, then
\begin{equation}
[p,q] = [l_1+1,l_2]
\label{1.3}
\end{equation}
The string susceptibility exponent $\gamma$ is
\begin{equation}
\gamma = \frac{-2}{p+q-1}
\label{1.4}
\end{equation}

The continuous series of critical points necessitates nonpolynomial critical
potentials that are holomorphic inside a circle of finite radius of convergence. 
Each depends on a parameter $l_1$ (respectively $l_2$). A third parameter,
a natural number n, is connected with the perturbative order at which the
equation
\begin{equation}
[B_2,B_1] = 1
\label{1.5}
\end{equation}
can be fulfilled and enters the string susceptibility exponent $\gamma$ as
\begin{equation}
\gamma = \frac{-2}{n-1}
\label{1.6}
\end{equation}

Whereas the discrete series is intimately connected with the theory of the
Korteweg-deVries equations \cite{5} and positive integer powers of quasi-differential
operators, we will use complex powers of such operators (as described in
\cite{6}) only marginally. The differential equations arising at the end
are trivial and have polynomial solutions.

The partition function $Z$ for the two-matrix model with action $S$
(\ref{1.1}) is
\begin{equation}
Z = \int \prod_{\alpha=1,2} \prod_{i \le j} d(Re M^{(\alpha)}_{ij})
\prod_{k < l} d(Im M_{kl}^{(\alpha)})e^{-S}
\label{1.7}
\end{equation}
The matrices $M^{(\alpha)}$ are diagonalized
\begin{equation}
M^{(\alpha)} = U^{(\alpha)} \Lambda^{(\alpha)} U^{(\alpha),-1}
\label{1.8}
\end{equation}
with unitary $N \times N$ matrices $U^{(\alpha)}$. After the integration
over these unitary matrices we have
\begin{equation}
Z = C(N) \int \prod_{\alpha=1,2} \prod_i d\lambda_i^{(\alpha)} \Delta
(\lambda^{(1)}) \Delta(\lambda^{(2)}) 
\cdot e^{S(\Lambda^{(1)}, \Lambda^{(2)})}
\label{1.9}
\end{equation}
$\quad (\Lambda^{(\alpha)} = {\rm diag} \{ \lambda_i^{(\alpha)}\}) $ \\
where $\Delta(\lambda)$ is the Vandermonde determinant
\begin{equation}
\Delta(\lambda) = \prod_{i<j}(\lambda_i-\lambda_j)
\label{1.10}
\end{equation}
The method of orthogonal polynomials \cite{7,8} applied to (\ref{1.9}) uses a
biorthogonal system
\begin{equation}
\{ \Pi_m(\lambda), \; \tilde{\Pi}_m(\mu)\}^{\infty}_{m=0}
\label{1.11}
\end{equation}
\begin{equation}
\deg \Pi_m = \deg \tilde{\Pi}_m = m
\label{1.12}
\end{equation}
so that
\begin{eqnarray}
& &\int\limits_{\bbbr_2} d\lambda d\mu \Pi_m (\lambda) \tilde{\Pi}_n(\nu) \nonumber \\
& &\times \exp \{ -V_1(\lambda) - V_2(\mu)+ c\lambda \mu \} = \delta_{nm}
\label{1.13}
\end{eqnarray}
Differentiation operators $A_1, A_2$ and multiplication operators
$B_1, B_2$ are introduced by 
\begin{eqnarray}
\Pi_m^{\prime}(\lambda) &=& \sum_n(A_1)_{mn} \Pi_n(\lambda) \nonumber \\
\tilde{\Pi}_m^{\prime}(\mu) &=& \sum_n(A_2)_{nm} \tilde{\Pi}_n(\mu)
\label{1.14}
\end{eqnarray}
and
\begin{eqnarray}
\lambda\Pi_m(\lambda) &=& \sum_n(B_1)_{mn} \Pi_n(\lambda) \nonumber \\
\mu \tilde{\Pi}_m(\mu) &=& \sum_n(B_2)_{nm} \tilde{\Pi}_n(\mu)
\label{1.15}
\end{eqnarray}
so that
\begin{equation}
[B_1,A_1] = [A_2,B_2] = 1
\label{1.16}
\end{equation}
We normalize $c$ in (\ref{1.13}) to one and derive Schwinger-Dyson equations
in the usual way
\begin{eqnarray}
A_1 + B_2 &=& V_1^{\prime}(B_1) \nonumber \\
A_2 + B_1 &=& V_2^{\prime}(B_2)
\label{1.17}
\end{eqnarray}
Then (\ref{1.16}) implies with (\ref{1.17})
\begin{equation}
[B_2,B_1] = 1
\label{1.18}
\end{equation}
From the definitions (\ref{1.14}), (\ref{1.15}) we deduce that
\begin{equation}
\begin{array}{rcl@{\qquad}c@{\qquad}rcl}
(A_1)_{mn} & = & 0 & \mbox{ except possibly for} & n-m & \le & -1 \\
(A_2)_{mn} & = & 0 & \mbox{ except possibly for} & n-m & \ge & +1 \\
(B_1)_{mn} & = & 0 & \mbox{ except possibly for} & n-m & \le & +1 \\
(B_1)_{mn} & = & 0 & \mbox{ except possibly for} & n-m & \ge & -1 
\end{array}
\label{1.19}
\end{equation}
If (\ref{1.17}) and (\ref{1.19}) are fulfilled, then the commutation
(\ref{1.18}) is diagonal. We call this assertion ``the lemma''. Contrary to
our procedure in \cite{4} we exploit (\ref{1.17}) only partially in this
work and therefore we impose (\ref{1.18}) as a strong additional constraint.

The Schwinger-Dyson equations (\ref{1.17}) are evaluated in the ``double
scaling limit''. All our notations are standard, in particular identical with
those in \cite{4}.

The critical potentials are expanded as
\begin{eqnarray}
V_1^{\prime}(t)^c &=& \sum^{\infty}_{k=0} f^c_k t^k \nonumber \\
V_2^{\prime}(t)^c &=& \sum^{\infty}_{k=0} g^c_k t^k
\label{1.20}
\end{eqnarray}
where $f^c_k (g^c_k)$ are certain analytic functions of $l_1$ and $l_2$.
We do not tune the coupling constants to these critical values but multiply
the whole critical action by a new parameter
\begin{equation}
\frac Ng
\label{1.21}
\end{equation}
Then we tune
\begin{equation}
N \to \infty, \quad g \to g_c
\label{1.22}
\end{equation}
The matrix labels $n, m$ become continuous in this limit
\begin{equation}
\frac nN = \xi, \quad 0 \le \xi \le 1
\label{1.23}
\end{equation}
We replace the label $N$ by the string coupling constant $a$
\begin{equation}
\frac1N = a^{2-\gamma}, \quad (\gamma < 0)
\label{1.24}
\end{equation}
so that $a \to 0$ if $N \to \infty$. Moreover
\begin{equation}
\xi = \frac{g_c}{g} (1-a^2x)
\label{1.25}
\end{equation}
We introduce a variable
\begin{equation}
z = e^{i\varphi}
\label{1.26}
\end{equation}
dual (in the Fourier series sense) to the discrete matrix label $n$. Then
we scale $\varphi$ as
\begin{equation}
\varphi = a^{-\gamma}p
\label{1.27}
\end{equation}
so that due to (\ref{1.24}), (\ref{1.25})
\begin{equation}
p = i \frac{d}{dx}
\label{1.28}
\end{equation}
is the quantum mechanical momentum operator corresponding to $x$.

New in this work is that the multiplication operator $B_1 (B_2)$ decomposes
into ``blocks'' corresponding to a nondegenerate $\bbbn^2$-lattice
\footnote{In this paper $\bbbn$ includes zero.}
\begin{equation}
B_1 = \sum_{[n_1,n_2]\in\bbbn^2} B_1^{[n_1,n_2]}
\label{1.29}
\end{equation}
where each block possesses a double scaling expansion
\begin{equation}
B_1^{[n_1,n_2]} \cong \sum^{\infty}_{n=0} a^{-\gamma[n+l_2-(n_1l_1+n_2l_2)]}
\cdot Q_n^{[n_1,n_2]}(x;p)
\label{1.30}
\end{equation}
Obviously it is necessary that
\begin{equation}
Re \, l_{1,2} < 0
\label{1.31}
\end{equation}
in order to render the expansion (\ref{1.29}) perturbative. Analogously we
expand $B_2$ in terms of $P_n^{[n_1,n_2]}(x;p)$. Both $Q_n$ and $P_n$ are
given as asymptotic expansions for $p \to + \infty$ (or $p \to - \infty$)
and with $p$ ordered to the right of $x$. They are quasidifferential
operators involving complex powers of the differential symbol $p$. For the
block $[0,0]$ which is ``basic'' in some sense, we found simple expressions
\begin{equation}
Q_0^{[0,0]} (x;p) = (e^{i\frac{\pi}{2}} L (x;p))^{l_2}  
\label{1.32} 
\end{equation}
\begin{equation}
P_0^{[0,0]} (x;p) = (e^{-i\frac{\pi}{2}} L (x;p))^{l_1}
\label{1.33}
\end{equation}
where $L$ has the form
\begin{equation}
L(x;p) = p + \sum^{\infty}_{n=1} u_n(x) p^{-n}
\label{1.34}
\end{equation}
The meaning of the complex labels $l_1$ and $l_2$ can be fixed by
(\ref{1.32}), (\ref{1.33}). The obvious commutativity of these operators
can be extended to the whole perturbative series (\ref{1.30})
\begin{equation}
[B_2^{[0,0]}, B_1^{[0,0]}] \cong 0
\label{1.35}
\end{equation}

\setcounter{equation}{0}
\section{The critical potentials} 
Our aim is now to solve the Schwinger-Dyson equations (\ref{1.17}), 
(\ref{1.19}) with analytic methods. We introduce the functions
\begin{equation}
b_1(z) = \sum^{+1}_{k=-\infty} (B_1)_{n,n+k}z^k
\label{2.1}
\end{equation}
\begin{equation}
b_2(z) = \sum^{+\infty}_{k=-1} (B_2)_{n,n+k}z^k
\label{2.2}
\end{equation}
where the n-dependence is not made explicit on the l.h.s. $b_1(z)\; (b_2(z))$
possesses a first order pole at infinity (zero) but is otherwise
assumed to be analytic in a punctured disc around infinity (zero). We fix
an irrelevant scale and assume that this disc extends to $z=1$. At $z=1$ 
either function will get a logarithmic branch point.

In polar coordinates we define the Fourier series boundary values
\begin{equation}
\lim_{r \searrow 1} b_1(re^{i\varphi}) = b_1(e^{i\varphi})_{\downarrow}
\label{2.3}
\end{equation}
\begin{equation}
\lim_{r \nearrow 1} b_2(re^{i\varphi}) = b_2(e^{i\varphi})_{\uparrow}
\label{2.4}
\end{equation}
Projection of the Fourier series $f(e^{i\varphi})$ on its non-negative 
(non-positive) frequency part is denoted $f(e^{i\varphi})_+ \;(f(e^{i\varphi})
_-)$. Then the Schwinger-Dyson equations (\ref{1.17}) with (\ref{1.19}) take the
form
\begin{equation}
(b_1)_{\downarrow-} = V_2^{\prime}(b_2)_{\uparrow-}
\label{2.5}
\end{equation}
\begin{equation}
(b_2)_{\uparrow+} = V_1^{\prime}(b_1)_{\downarrow+}
\label{2.6}
\end{equation}
In the double-scaling asymptotic region we expand the functions $b_{1,2}(z)$
(\ref{2.1}), (\ref{2.2}) as
\begin{equation}
b_1(z) = r(z) + \sum^{\infty}_{m=1} a^{-(m+1)\gamma} U_m(x;z)
\label{2.7}
\end{equation}
\begin{equation}
b_2(z) = s(z) + \sum^{\infty}_{m=1} a^{-(m+1)\gamma} V_m(x;z)
\label{2.8}
\end{equation}
where the functions $U_m \; (V_m)$ are later expanded in the neighborhood
of $z = 1$ into an asymptotic power series in $(1-\frac1z)$ (or $(1-z)$).
For $r(z)$ and $s(z)$ we make the ansatz ($l_1, l_2 \in \bbbc$, (\ref{1.31}))
\begin{eqnarray}
\label{2.9}
r(z) &=& z(1-\frac1z)^{l_2} = \sum^{\infty}_{k=-1} \rho_{-k}z^{-k} \\
& &\quad (|z| > 1) \nonumber
\end{eqnarray}
\begin{eqnarray}
\label{2.10}
s(z) &=& \frac1z(1-z)^{l_1} = \sum^{\infty}_{k=-1} \sigma_k z^k \\
& &\quad (|z| < 1) \nonumber
\end{eqnarray}
At leading order (\ref{2.6}) goes into
\begin{equation}
k \ge 0: \; \sigma_k = \sum^{\infty}_{s=0} M(l_2)_{ks}\, f^c_s
\label{2.11}
\end{equation}
where $\{f^c_s\}^{\infty}_0$ are the critical coupling constants of the
potential $V_1$ (see (\ref{1.20})) and from (\ref{2.10})
\begin{equation}
M(l)_{ks} = (-1)^{s-k} {ls \choose s-k}
\label{2.12}
\end{equation}
This matrix $M(l)$ is upper triangular with ones on the diagonal. So its
inverse exists, but we must know whether $\{\sigma_k\}$ is in the domain
of $M(l)^{-1}$ when we invert (\ref{2.11}).

In order to invert $M(l)$ we invert the function
\begin{equation}
t = r(z)
\label{2.13}
\end{equation}
to
\begin{equation}
z = \beta(t)
\label{2.14}
\end{equation}
so that for $z \to \infty$
\begin{equation}
\beta(t) = t + l_2 - {l_2 \choose 2} \frac1t + O\left( \frac{1}{t^2} \right)
\label{2.15}
\end{equation}
This function $\beta(t)$ is holomorphic for
\begin{eqnarray}
|t| > |t_c| : \nonumber \\
t_c = (-l_2)^{l_2} (1-l_2)^{1-l_2}
\label{2.16}
\end{eqnarray}
where the principal branch of the logarithm with cut on the negative real axis
is used. When $t$ approaches $t_c$
\begin{equation}
\beta(t) \sim A\left( 1 - \frac{t_c}{t} \right) ^{\frac12}
\label{2.17}
\end{equation}
This singularity arises from a stationary point of $r(z)$, and (\ref{2.16})
can be obtained this way.

If we define
\begin{equation}
\beta(t)^n = \sum^n_{k=-\infty} a_{kn} t^k
\label{2.18}
\end{equation}
we obtain
\begin{equation}
a_{kn} = \left\{ \begin{array}{lr}
(-1)^{n-k} {-kl_2 \choose n-k} \frac nk & (k \not= 0) \\
\delta_{n0} + l_2(1- \delta_{n0}) & (k=0)
\end{array} \right.
\label{2.19}
\end{equation}
The expansion coefficients in (\ref{2.15}) are $a_{k1}$, and we will see in 
a moment that
\begin{equation}
(M(l_2)^{-1})_{mn} = a_{mn}
\label{2.20}
\end{equation}
To prove this we note that for $z \to \infty$ (\ref{2.13}), (\ref{2.15})
imply
\begin{equation}
(r(z)^k)_+ = t^k + O\left( \frac1t \right)
\label{2.21}
\end{equation}
and consequently
\begin{equation}
V_1^{\prime}(r(z))_+ = V_1^{\prime}(t) +  O\left( \frac1t \right)
\label{2.22}
\end{equation}
We will see in a moment that $V_1^{\prime}(t)$ converges for (with 
(\ref{2.16}))
\[ |t| < |t_c|  \]
so that for $t \to t_c$
\begin{equation}
V_1^{\prime}(t) \sim C \left(1-\frac{t}{t_c} \right)^{\frac12}
\label{2.23}
\end{equation}
since the exponents in (\ref{2.17}), (\ref{2.23}) are bigger than the limit
of integrability -1, (see our discussion in the Appendix) the Fourier series 
obtained in the limit
\begin{equation}
t = |t_c| e^{i\Theta}
\label{2.24}
\end{equation}
are equal
\begin{equation}
V_1^{\prime} (|t_c|e^{i\Theta})_{\uparrow,+} = \sum^{\infty}_{k=0}
\sigma_k (\beta(|t_c|e^{i\Theta})^k)_{\downarrow,+}
\label{2.25}
\end{equation}
provided the sum on the r.h.s. converges appropriately. From (\ref{2.25}) 
follows (\ref{2.20}) immediately.

The critical coupling constants can be obtained from (\ref{2.11}), (\ref{2.20})
and (\ref{2.19})
\begin{eqnarray}
f^c_k &=& \sum^{\infty}_{m=0} a_{km} \sigma_m \nonumber \\
&=& (l_1l_2-l_1-l_2) \frac{(-1)^k \Gamma(l_1-l_2k)}{\Gamma(l_1-k) \Gamma
(2+(1-l_2)k)}
\label{2.26}
\end{eqnarray}
Since the summation in (\ref{2.26}) is hypergeometric, absolute convergence
follows from
\begin{equation}
Re (l_1-l_2k) > 0
\label{2.27}
\end{equation}
Due to (\ref{1.3}) this may be violated for a finite number of k. In these
cases we postulate (\ref{2.26}) to be true by ``analytic continuation''. 
Of course this amounts to a renormalization by subtraction of infinite 
counter terms. Using (\ref{2.26}) we can by applying Stirling's formula
immediately prove (\ref{2.23}).
\setcounter{equation}{0}
\section{Evaluation of the Schwinger-Dyson equations}
The Schwinger-Dyson equations (\ref{1.17}) are evaluated in the neighborhood
of $z=1$ by an asymptotic expansion in powers of $1-z$ or $1-\frac1z$. The 
functions $U_m(x;z), V_m(x;z)$ in (\ref{2.7}), (\ref{2.8}) are decomposed
into contributions of blocks, too
\begin{eqnarray}
U_m(x;z) &=& \sum_{[n_1,n_2]\in\bbbn^2} \;\sum^{\infty}_{r=0}
U^{[n_1,n_2]}_{mr}(x) \nonumber \\
& & \times z\left(1-\frac1z\right)^{l_2-(n_1l_1+n_2l_2)-(m+1)+r}
\label{3.1}
\end{eqnarray}
\begin{eqnarray}
V_m(x;z) &=& \sum_{[n_1,n_2]\in\bbbn^2} \;\sum^{\infty}_{r=0}
V^{[n_1,n_2]}_{mr}(x) \nonumber \\
& & \times \frac1z (1-z)^{l_1-(n_1l_1+n_2l_2)-(m+1)+r}
\label{3.2}
\end{eqnarray}
The appearance of these blocks is a necessary consequence of the recursion
relations that we will derive next. Otherwise the form of the expansion is
an intuitive generalisation of what has been found for the discrete
series \cite{4}.

These recursion relations are derived from the identities
\begin{equation}
\left( \frac1z(1-z)^{l_1} \right)_{\uparrow,+} = \sum^{\infty}_{k=0}
f^c_k \left( z^k(1-\frac1z)^{kl_2} \right)_{\downarrow,+} 
\label{3.3}
\end{equation}
\begin{equation}
\left(z(1-\frac1z)^{l_2} \right)_{\downarrow,-} = \sum^{\infty}_{k=0}
g^c_k \left( z^{-k}(1-z)^{kl_1} \right)_{\uparrow,-} 
\label{3.4}
\end{equation}
from which we derived the critical coupling constants (\ref{2.26}). We 
shall denote (\ref{3.4}) ``dual'' to (\ref{3.3}) and extend this notation
also on the recursion relations derived from (\ref{3.4}). This ``duality''
is connected with the replacements
\begin{equation}
l_1 \longleftrightarrow l_2, \; z \longleftrightarrow \frac1z
\label{3.5}
\end{equation}
On the r.h.s. of (\ref{3.3}) we have terms (see (\ref{2.9}))
\[ f^c_k r(z)^k \]
If we replace one factor $r(z)$ by an order-m term from (\ref{2.7}) and
use (\ref{3.1}) we obtain
\begin{eqnarray}
\label{3.6}
U^{[n_1,n_2]}_{ms} (x) \sum^{\infty}_{k=1} f^c_k k z^k \left(1-\frac1z \right)
^{kl_2-\lambda-(m+1)+s} \nonumber \\
(\lambda = n_1l_1 + n_2l_2)
\end{eqnarray}
We make an ansatz 
\begin{eqnarray}
& &\left( \sum^{\infty}_{k=1} k f^c_k z^k \left(1-\frac1z \right)^{kl_2}\right)
_{\downarrow,+} = \nonumber \\
& &= \left( \sum^{\infty}_{n=0} c_n (1-z)^{l_1+n}\right)
_{\uparrow,+} \nonumber \\
& &+ \mbox{ holomorphic function at } z=1
\label{3.7}
\end{eqnarray}
Taking derivatives $z \frac{d}{dz}$ of (\ref{3.3}) (taking derivatives is
compatible with the operations $\downarrow,+$ etc.) gives
\begin{equation}
c_n = + \frac{l_1 + \sum^n_{k=1} l^k_2}{l_2^{n+1}}
\label{3.8}
\end{equation}
The holomorphic part in (\ref{3.7}) is unavoidable and is essential for the
sequel, remember that in the case of the discrete series there are 
\underline{only} holomorphic parts.

Inserting (\ref{3.7}) into (\ref{3.6}) we obtain
\[ \left( (1-\frac1z)^{-\lambda-(m+1)+s} \right)_{\downarrow}
\left((1-z)^{l_1+n} \right)_{\uparrow} \]
which we evaluate with the help of the identity $(\epsilon \searrow 0)$
\begin{eqnarray}
& &( (1-e^{-i\varphi-\epsilon})^a (1-e^{+i\varphi-\epsilon})^b )_+ 
\nonumber \\
& &= \frac{\sin \pi b}{\sin\pi(a+b)} (1-e^{i\varphi-\epsilon})^{a+b}
(1-(1-e^{i\varphi-\epsilon}))^{-a} \nonumber \\
& &+ {a+b-1 \choose b} \;_2F_1(1,-b;1-a-b;1-e^{i\varphi-\epsilon})
\label{3.9}
\end{eqnarray}
which is valid for $\varphi \to 0, a+b \notin \bbbz$. We have made the 
holomorphic part in (\ref{3.9}) explicit. If $b \in \bbbz, a+b \notin \bbbz$,
only this holomorphic part of (\ref{3.9}) survives. Therefore the holomorphic
function in (\ref{3.7}) contributes only to the holomorphic part of
(\ref{3.9}). 

The recursion relations that can be derived in this fashion express
functions
\[ V_{mr}^{[n_1,n_2]}(x), \quad [n_1,n_2] \not= [1,0] \]
in terms of the functions $U_{m^{\prime}s}^{[n_1^{\prime},n_2^{\prime}]}(x)$ 
and their derivatives. First there is a linear contribution $(\lambda =
n_1l_1 + n_2l_2)$
\begin{eqnarray}
&+& \frac{\sin \pi l_1}{\sin \pi(l_1-\lambda)} \sum^r_{n=0} \sum^{r-n}_{s=0}
(-1)^{m+n+r+1} \nonumber \\
& & \times {\lambda + m + 2 - s \choose r-n-s} c_n U_{ms}^{[n_1,n_2]}(x)
\label{3.10}
\end{eqnarray}
In order to obtain expressions which are quadratic in $U$ or linear in its 
derivatives we need sets of coefficients generalizing the $\{c_n\}$ in 
(\ref{3.7})
\begin{eqnarray}
\left( \sum^{\infty}_{k=1} k^2 f^c_k z^k \left(1-\frac1z \right)^{kl_2} \right)
_{\downarrow,+} \nonumber \\
= \left( \sum^{\infty}_{n=0} d_n (1-z)^{l_1+n} \right)_{\uparrow,+}\\
+ \mbox{ holomorphic function at } z = 1 \nonumber
\label{3.11}
\end{eqnarray}
\begin{eqnarray}
\left( \sum^{\infty}_{k=1} k^3 f^c_k z^k \left(1-\frac1z \right)^{kl_2} \right)
_{\downarrow,+} \nonumber \\
= \left( \sum^{\infty}_{n=0} e_n (1-z)^{l_1+n} \right)_{\uparrow,+}\\
+ \mbox{ holomorphic function at } z = 1 \nonumber
\label{3.12}
\end{eqnarray}
By differentiating (\ref{3.3}) twice or three times we obtain the recursions
\begin{eqnarray}
& &l^2_2 d_n - 2l_2d_{n-1} + d_{n-2} = + l_2(c_n-c_{n-1}) + 1 \nonumber \\
& &+ (l^2_1-l_1-1) \delta_{n0} - (l_1-1)^2 \delta_{n1} \nonumber \\
& &(d_n = 0 \;{\rm for} \; n < 0)
\label{3.13}
\end{eqnarray}
\begin{eqnarray}
& &-l^3_2 e_n + 3l^2_2 e_{n-1} - 3l_2e_{n-2} + e_{n-3} \nonumber \\
& &+ 3l_2 (l_2d_n - (l_2+1)d_{n-1} + d_{n-2}) \nonumber \\
& &-l_2 (2c_n - 3c_{n-1} + c_{n-2}) = \nonumber \\
& &-1 -[l_1(l_1-1)(l_1-2)-1] \delta_{n0} \nonumber \\
& &+ [2l_1(l_1-1) (l_1-2) + 1] \delta_{n1} - (l_1-1)^3 \delta_{n2} \nonumber \\
& &(e_n = 0 \;{\rm for} \; n < 0) 
\label{3.14}
\end{eqnarray}
from which we derive that 
\begin{equation}
d_0 = + \frac{l^2_1}{l^2_2}
\label{3.15}
\end{equation}
\begin{equation}
d_1 = \frac{l_1(2l_1+1)}{l^3_2} - \frac{l^2_1-l_1-1}{l^2_2}
\label{3.16}
\end{equation}
\begin{equation}
e_0 = + \frac{l^3_1}{l^3_2}
\label{3.17}
\end{equation}
\begin{equation}
e_1 = + \frac{l_1}{l^4_2} (3l^2_1 + 3l_1 + 1) - \frac{1}{l^3_2}
(2l^3_1 - 2l_1 -1)
\label{3.18}
\end{equation}
With the help of these sets of coefficients we obtain as further contributions 
to the function $V_{mr}^{[n_1,n_2]}(x)$ 
\begin{eqnarray}
& &+ \frac{\sin \pi l_1}{\sin \pi(l_1-\lambda)} \sum_n \big\{
\sum_{s_1,s_2} (-1)^{m+n+r+1} {\lambda+m+2-s_1-s_2 \choose r-n-s_1-s_2}
\frac12 (d_n-c_n) \nonumber \\
& &\times \sum_{n_1^{\prime},n_2^{\prime},m_1} U_{m_1s_1}^{[n_1^{\prime},n_2^{\prime}]}
(x) U_{m-m_1-1,s_2}^{[n_1-n_1^{\prime},n_2-n_2^{\prime}]} (x) \nonumber \\
& &+ \sum_{s=0}^{r-n} (-1)^{m+n+r} {\lambda+m+1-s \choose r-n-s} \frac12
[l_2(d_n-c_n)-(d_{n-1}-c_{n-1})] \frac{d}{dx} U_{m-1,s}^{[n_1,n_2]}(x)
\nonumber \\
& &+ \sum_{s=0}^{r-n} (-1)^{m+n+r+1} {\lambda+m-s \choose r-n-s} \frac{1}{12}
[2(l^2_2e_n-2l_2e_{n-1}+e_{n-2}) \nonumber \\
& &- 3(l_2(l_2+1)d_n-3l_2d_{n-1}+d_{n-2})
\nonumber \\
& &+ l_2(l_2+3)c_n-5l_2c_{n-1}+c_{n-2}] \frac{d^2}{dx^2} U_{m-2,s}^{[n_1,n_2]}
(x) \nonumber \\
& &+ O(U^3,UU^{\prime}, UU^{\prime\prime}, (U^{\prime})^2, U^{\prime\prime
\prime},...) \big\}
\label{3.19}
\end{eqnarray}
The corresponding dual relations are obtained from (\ref{3.4}) and express
\[ U_{mr}^{[n_1,n_2]} (x) \]
through $V_{m^{\prime},s}^{[n_1^{\prime},n_2^{\prime}]}(x)$ by means of
coefficients $\{\tilde{c}_n\},\{\tilde{d}_n\}, \{\tilde{e}_n\}\ldots$ which
are derived from $\{ c_n\}, \{d_n\}, \{e_n\} \ldots$ by the replacement
$l_1 \longleftrightarrow l_2$. Before we are able to show that the full set of 
relations is ``recursive'' in the proper sense, we have to eliminate ``circles''.

First we mention that there are two types of gradings which are ``conserved''
by these relations.
\begin{enumerate}
\item[($\alpha$)] 
We define the ``block-grade''of
\[ \left( \frac{d}{dx} \right)^k U_{mr}^{[n_1,n_2]} (x) \]
to be
\[ [n_1,n_2] \]
Then $V_{m,s}^{[n_1,n_2]}(x)$ obtains contributions of only those monomials in
$ \left( \frac{d}{dx} \right)^{k_i} \newline U_{m_i,r_i}^{[n_{1i},n_{2i}]} (x)$ 
so that
\begin{eqnarray}
n_1 &=& \sum_i n_{1i} \nonumber \\
n_2 &=& \sum_i n_{2i}
\label{3.20}
\end{eqnarray}
\item[($\beta$)]
We define the ``m-grade'' of
\[ \left( \frac{d}{dx} \right)^k U_{mr}^{[n_1,n_2]} (x) \]
to be 
\begin{equation}
\mu = m + 1+k
\label{3.21}
\end{equation}
Then $V_{m,s}^{[n_1,n_2]}(x)$ obtains contributions of only those monomials in
$ \left( \frac{d}{dx} \right)^{k_i} \\U_{m_i,r_i}^{[n_{1i},n_{2i}]} (x)$ 
so that
\begin{equation}
m+1 = \sum _i \mu_i
\label{3.22}
\end{equation}
The recursion of (\ref{3.10}), (\ref{3.19}) and their duals is in both grades.
\end{enumerate}

We denote the blocks
\[ [0,0], \; [0,1], \; [1,0], \; [1,1] \]
the ``elementary blocks''. The blocks $[n_1,n_2]$ with $n_1 > 1$ or
$n_2 > 1$ are called ``nonelementary''. We show first that we can express
recursively
\[ U_{mr}^{[n_1,n_2]}(x), \; V_{m^{\prime},s}^{[n_1^{\prime},n_2^{\prime}]}(x)
\]
from nonelementary blocks by the functions of the elementary blocks. Namely 
reinsert the dual recursion relation for $U_{ms}^{[n_1,n_2]}$ into the
recursion relation for $V_{mr}^{[n_1,n_2]}$
\begin{eqnarray}
V_{mr}^{[n_1,n_2]}(x) &=& \frac{\sin \pi l_1 \sin \pi l_2}
{\sin\pi (l_1-\lambda) \sin \pi(l_2-\lambda)} \sum^r_{s=0}
A_{rs} V_{ms}^{[n_1,n_2]}(x) \nonumber \\
&+& \mbox{ derivative and nonlinear terms}
\label{3.23}
\end{eqnarray}
where
\begin{eqnarray}
A_{rs} &=& \sum^r_{s^{\prime}=s} \,\sum^{r-s^{\prime}}_{n=0} \,\sum
^{s^{\prime}-s}_{n^{\prime}=0} \, (-1)^{n+n^{\prime}+r+s^{\prime}} \nonumber \\
& & {\lambda + m+2-s^{\prime} \choose r-n-s^{\prime}}
{\lambda + m+2-s \choose s^{\prime}-n^{\prime}-s} c_n \tilde{c}_{n^{\prime}}
\label{3.24}
\end{eqnarray}
This matrix can be shown to be the unit matrix. For
\[ \lambda = n_1l_1 + n_2l_2 \]
not belonging to an elementary block we have
\begin{equation}
\frac{\sin\pi l_1 \sin\pi l_2}{\sin\pi(l_1-\lambda) \sin \pi(l_2-\lambda)}
\not= 1
\label{3.25}
\end{equation}
Thus we cast the first term of the r.h.s. of (\ref{3.23}) on the l.h.s. and 
divide by
\begin{equation}
1 - \frac{\sin\pi l_1 \sin\pi l_2}{\sin\pi(l_1-\lambda) \sin \pi(l_2-\lambda)}
\label{3.26}
\end{equation}

Next we come to the elementary blocks which we deal with one after the
other. The basic block $[0,0]$ is also the simplest one. The recursion
relations for $U_{mr}^{[0,0]}(x), V_{mr}^{[0,0]}(x)$ involve only
functions of the same block $[0,0]$ and are therefore recursive only in
the m-grade. The connection of the recursion relations with the dual 
recursion relations is quite simple: one set is the inverse of the other
set. So every function $V_{mr}^{[0,0]}(x)$ can be expressed as a polynomial
in $U_{m^{\prime}s}^{[0,0]}(x)$ and its derivatives, and these
$U_{m^{\prime}s}^{[0,0]}(x)$ can be considered as being free.

Next we study the block $[1,0]$. From the ansatz (\ref{3.2}) we see that the
corresponding parts of the functions $V_m(x;z)$ are single-valued in a
neighborhood of $z=1$, i.e. they have only zeros and poles. From (\ref{3.9})
we recognize that poles are in fact not needed. Therefore we make the 
``no-pole-assumption''
\begin{equation}
V_{mr}^{[1,0]}(x) = 0 \; \mbox{for} \; m \ge r
\label{3.27}
\end{equation}
One can also show easily that if the Schwinger-Dyson equations are 
evaluated asymptotically for $z \to 1$, pole terms are unconstrained.

By means of the dual recursion relations we can express the 
$U_{mr}^{[1,0]}(x)$ as polynomials in $V_{mr}^{[1,0]}(x), 
\; V_{mr}^{[0,0]}(x)$ and their derivatives. Application of the direct
recursion makes no sense, because the holomorphic part of the functions
$V_m(x;z)$ is produced by a completely different mechanism. They stem
from an infinite number of sources and cannot be computed. We shall
therefore consider these functions as freely eligible.

The block $[0,1]$ behaves very similar as the block $[1,0]$. The
``no-pole-assump- tion'' is
\begin{equation}
U_{mr}^{[0,1]}(x) = 0 \, \mbox{for} \, m \ge r
\label{3.28}
\end{equation}
The functions $V_{mr}^{[0,1]}(x)$ are expressed as polynomials in
$U_{mr}^{[0,1]}(x), \; U_{mr}^{[0,0]}(x)$ and their derivatives by the 
recursion relations. The dual recursion relations are suppressed.

Finally we come to the block $[1,1]$. Of course (\ref{3.26}) is zero. But the
matrix A in (\ref{3.24}) is also the unit matrix. Thus the l.h.s. of
(\ref{3.23}) cancels the first term of the r.h.s. What is the remainder?
Taking into account all recursion relations for the blocks $[0,0], \,
[0,1], \, [1,0]$, we can show that the remainder vanishes, too. Therefore the
recursion relations and dual recursion relations are inverses of each other
modulo the recursion relations of the other elementary blocks.

Thus we have obtained an algorithm by which all functions
\[ U_{mr}^{[n_1,n_2]}(x), \; V_{mr}^{[n_1,n_2]}(x) \]
can be expressed as polynomials in
\begin{eqnarray}
& &U_{mr}^{[0,0]}(x), \; V_{mr}^{[1,0]}(x) \; (r \ge m+1), \nonumber \\
& &U_{mr}^{[0,1]}(x) \; (r \ge m+1), \; U_{mr}^{[1,1]}(x)\\
& &(m \ge +1 \mbox{ in all cases}) \nonumber 
\label{3.29}
\end{eqnarray}
and their derivatives. It is not possible to restrict the blocks from 
$\bbbn^2$ to the single block $[0,0]$. The $\bbbn^2$ lattice is generated by
the functions $V^{[1,0]}_{mr}(x), U^{[0,1]}_{mr}(x)$ that can always be
present on the r.h.s. of the recursion relations and their duals and 
correspond to holomorphic behaviour at $z=1$.

\setcounter{equation}{0}
\section{The double scaling limits of $B_1$ and $B_2$}
In the double scaling limit the $a^{-\gamma}$-expansions (\ref{2.7}),
(\ref{2.8}) are combined with the $(1-z)$-expansions (\ref{3.1}),
(\ref{3.2}), and (\ref{1.26}), (\ref{1.27}) are inserted so that a simple 
expansion in powers of $a^{-\gamma}$ results:
\begin{eqnarray}
B_1 &=& \sum_{[n_1,n_2]\in\bbbn^2} \; B_1^{[n_1,n_2]} \\
\label{4.1}
B_2 &=& \sum_{[n_1,n_2]\in\bbbn^2} \; B_2^{[n_1,n_2]} 
\label{4.2}
\end{eqnarray}
and with 
\begin{equation}
\lambda = n_1l_1 + n_2l_2
\label{4.3}
\end{equation}
\begin{equation}
B_1^{[n_1,n_2]} \simeq \sum^{\infty}_{n=0} a^{-(l_2-\lambda+n)\gamma}
Q_n^{[n_1,n_2]}(x;p)
\label{4.4}
\end{equation}
\begin{equation}
B_2^{[n_1,n_2]} \simeq \sum^{\infty}_{n=0} a^{-(l_1-\lambda+n)\gamma}
P_n^{[n_1,n_2]}(x;p)
\label{4.5}
\end{equation}
The $P_n^{[n_1,n_2]}, \, Q_n^{[n_1,n_2]}$ are quasidifferential operators.
The contributions of $r(z)$ (\ref{2.7}) and $s(z)$ (\ref{2.8}) are
attributed to the block $[0,0]$.

We define expansion coefficients
\begin{equation}
e^x(1-e^{-x})^l = \sum^{\infty}_{k=0} t_k(l) x^{k+l}
\label{4.6}
\end{equation}
so that the first few are
\begin{eqnarray}
t_0(l) &=& 1 \nonumber \\
t_1(l) &=& 1 - \frac12 {l \choose 1} \nonumber \\
t_2(l) &=& \frac12 - \frac13 {l \choose 1} + \frac14 {l \choose 2}
\label{4.7}
\end{eqnarray}
From the $U$'s and $V$'s we go over to new functions $(\lambda = n_1l_1 
+ n_2l_2)$
\begin{equation}
\Phi_{mn}^{[n_1,n_2]}(x) = \sum^n_{r=0} t_{n-r} (l_2-\lambda - (m+1) + r)
 U_{mr}^{[n_1,n_2]}(x)
\label{4.8}
\end{equation}
\begin{equation}
\Psi_{mn}^{[n_1,n_2]}(x) = \sum^n_{r=0} t_{n-r} (l_1-\lambda - (m+1) + r)
 V_{mr}^{[n_1,n_2]}(x)
\label{4.9}
\end{equation}
$\quad (1 \le m < \infty)$ \\
and
\begin{equation}
\Phi_{-1,n}^{[0,0]} = t_n(l_2)
\label{4.10}
\end{equation}
\begin{equation}
\Psi_{-1,n}^{[0,0]} = t_n(l_1)
\label{4.11}
\end{equation}
The quasidifferential operators $Q_n^{[n_1,n_2]}, \, P_n^{[n_1,n_2]}$
are then given by
\begin{equation}
Q_n^{[n_1,n_2]}(x;p) = \sum_m \Phi_{mn}^{[n_1,n_2]}(x) \left(e^{i\frac
{\pi}{2}}p \right)^{-n_1l_1-(n_2-1)l_2-(m+1)+n}
\label{4.12}
\end{equation}
\begin{equation}
P_n^{[n_1,n_2]}(x;p) = \sum_m \Psi_{mn}^{[n_1,n_2]}(x) \left(e^{-i\frac
{\pi}{2}}p \right)^{-(n_1-1)l_1-n_2l_2-(m+1)+n}
\label{4.13}
\end{equation}
Here the summation over $m$ extends from $\min m$ to infinity where
\begin{equation}
\min (m+1) = 2 \max (n_1,n_2)
\label{4.14}
\end{equation}
as a consequence of the recursion relations (\ref{3.10}), (\ref{3.19}).

Now we face the problem to solve all the constraints following from the
commutator (\ref{1.5}), (\ref{1.18}). Since
\begin{eqnarray}
[B_2^{[n_1,n_2]}, B_1^{[n_1^{\prime},n_2^{\prime}]}] \cong \sum^{\infty}_{n=0} 
a^{-\gamma[n-(n_1+n_1^{\prime}-1)l_1-(n_2+n_2^{\prime}-1)l_2]} \nonumber \\
\times \sum^n_{s=0} [P_{n-s}^{[n_1,n_2]}, Q_s^{[n_1^{\prime},n_2^{\prime}]}]
\label{4.15}
\end{eqnarray}
and
\begin{eqnarray}
[P_{n-s}^{[n_1,n_2]}, Q_s^{[n_1^{\prime},n_2^{\prime}]}] = \hspace{9.5cm}\nonumber \\
= \sum^{\infty}_{k=1} \sum_{m_1,m_2} \exp i \frac{\pi}{2}
[2s-n+(n_1-n_1^{\prime}-1)l_1+(n_2-n_2^{\prime}+1)l_2+m_1-m_2] \nonumber \\
\times \Big\{ {l_1-\lambda+n-s-m_1-1 \choose k} \Psi^{[n_1,n_2]}
_{m_1,n-s} (x) \left( i \frac{d}{dx} \right)^k \Phi_{m_2,s}^{[n_1^{\prime},
n_2^{\prime}]}(x) \nonumber \\
- {l_2-\lambda^{\prime}+s-m_2-1 \choose k} \Phi^{[n_1^{\prime},n_2^{\prime}]}
_{m_2,s} (x) \left( i \frac{d}{dx} \right)^k \Psi_{m_1,n-s}^{[n_1,n_2]}(x)
\Big\} \nonumber \\
\times p^{n-(n_1+n_1^{\prime}-1)l_1 - (n_2+n_2^{\prime}-1)l_2 - m_1-m_2-k-2}
\hspace{3.5cm}
\label{4.16}
\end{eqnarray}
Integer powers of $p$ result only if (for generic $l_1$ and $l_2$)
\begin{eqnarray}
n_1 + n_1^{\prime} = 1 \nonumber \\
n_2 + n_2^{\prime} = 1
\label{4.17}
\end{eqnarray}
All commutators (\ref{4.15}) not satisfying (\ref{4.17}) must vanish,
whereas
\begin{equation}
\sum_{\begin{array}{l}
n_1,n_1^{\prime},n_2,n_2^{\prime} \\ n_1+n_1^{\prime} = 1 \\
n_2+n_2^{\prime} = 1 \end{array}}
[B_2^{[n_1,n_2]}, B_1^{[n_1^{\prime},n_2^{\prime}]}] = 1
\label{4.18}
\end{equation}
We shall now demonstrate that this problem can be solved with the sole free
functions (\ref{3.29}).

We assume first that for all $m,r$
\begin{equation}
\frac{d}{dx} U_{mr}^{[0,0]} = \frac{d}{dx} U_{mr}^{[0,1]} =
\frac{d}{dx} V_{mr}^{[1,0]} = 0 
\label{4.19}
\end{equation}
Then of all the commutators (\ref{4.16}) those with
\begin{equation}
n_1 \cdot n_2 = 0 \; {\rm and} \; n_1^{\prime} \cdot n_2^{\prime} = 0
\label{4.20}
\end{equation}
vanish trivially whereas (\ref{4.18}) reduces to
\begin{equation}
[B_2^{[0,0]}, B_1^{[1,1]}] + [B_2^{[1,1]}, B_1^{[0,0]}] = 1
\label{4.21}
\end{equation}
Next assume that $n_0 \ge 3$ exists to that 
\begin{equation}
\frac{d}{dx} U_{mn}^{[1,1]} = 0 \,\mbox
{for all}\, m \, \mbox{and}\, n < n_0
\label{4.22}
\end{equation}
Then for the leading order
\begin{equation}
a^{-\gamma n_0}
\label{4.23}
\end{equation}
in (\ref{4.15}) we obtain from (\ref{4.21}) using (\ref{4.16}) and $k=1$
\begin{eqnarray}
\sum^{n_0}_{s=0} \sum_{m_1,m_2} & & \Big\{ e^{i \frac{\pi}{2}(2s-n_0-2l_1
+m_1-m_2)} \nonumber \\
& &\times (l_1+n_0-s-m_1-1) \Psi_{m_1,n_0-s}^{[0,0]}(x) i \frac{d}{dx}
\Phi_{m_2,s}^{[1,1]} \nonumber \\
& & - e^{i \frac{\pi}{2}(2s-n_0+2l_2+m_1-m_2)} \nonumber \\
& & \times(l_2+s-m_2-1) \Phi^{[0,0]}_{m_2,s}(x) i \frac{d}{dx}
\Psi^{[1,1]}_{m_1,n_0-s}(x) \big\} \nonumber \\
& & \times p^{n_0-m_1-m_2-3}
\label{4.24}
\end{eqnarray}
so
\begin{equation}
n_0 = 3
\label{4.25}
\end{equation}
is possible for
\begin{eqnarray}
m_1 = -1, \quad m_2 = +1, \quad \mbox{(first term)} \nonumber \\
m_1 = +1, \quad m_2 = -1, \quad \mbox{(second term)}
\label{4.26}
\end{eqnarray}
In this case we obtain
\begin{equation}
\left[ - l_1e^{-i\pi l_1} - l_1 e^{+i\pi l_2} \cdot \frac{\sin \pi l_1}
{\sin \pi l_2} \right] \frac{d}{dx} U_{13}^{[1,1]}(x) = 1
\label{4.27}
\end{equation}
The bracket is
\begin{equation}
- l_1 \frac{\sin \pi (l_1+l_2)}{\sin \pi l_2} \not= 0
\label{4.28}
\end{equation}
and after normalization of $x$ by translation we obtain
\begin{equation}
U_{13}^{[1,1]}(x) = \alpha_{13}x
\label{4.29}
\end{equation}
Letting $m_1$ and $m_2$ grow beyond the values (\ref{4.26}) we obtain
\begin{equation} 
U_{m3}^{[1,1]}(x) = \alpha_{m3}x + \beta_{m3}
\label{4.30}
\end{equation}
where all $\alpha_{m3}$ are determined (e.g. $\alpha_{23} = 0$) but all
$\beta_{m3}$ are free integration constants (except $\beta_{13} = 0$).

Choosing $n_0 > 3$ we can show in the same way that
\begin{equation}
U_{mn_0}^{[1,1]}(x) = \alpha_{mn_0}x + \beta_{mn_0}
\label{4.31}
\end{equation}
Given any $n_0 \ge 3 \;\gamma$ is fixed by the usual argument to
\begin{equation}
\gamma = \frac{-2}{n_0-1}
\label{4.32}
\end{equation}

Still all commutator constraints must be fulfilled where
\begin{equation}
\max (n_1+n_1^{\prime}-1, \, n_2+n_2^{\prime}-1) > 0
\label{4.33}
\end{equation}
But this is necessary only up to (and including) the order (\ref{4.23}), 
namely whenever (see (\ref{4.25}))
\begin{equation}
n - (n_1+n_1^{\prime}-1) Re l_1 - (n_2+n_2^{\prime}-1) Re l_2 \le n_0
\label{4.34}
\end{equation}
Now among the two functions
\begin{equation}
\Phi_{m_2,s}^{[n_1^{\prime},n_2^{\prime}]}(x), \quad
\Psi_{m_1,n-s}^{[n_1,n_2]}(x)
\label{4.35}
\end{equation}
at least one must contain a function
\[ U_{m,r}^{[1,1]}, \; r \ge n_0 \]
which is nonconstant if (\ref{4.16}) is nonzero. Let us assume that this is 
the first in (\ref{4.35}), so that
\[ n_1^{\prime} \ge 1, \; n_2^{\prime} \ge 1 \]
Then from (\ref{4.34}) together with (\ref{4.33}) we obtain
\begin{equation}
n < n_0
\label{4.36}
\end{equation}
On the other hand from (\ref{4.35})
\begin{equation}
s \ge r \ge n_0
\label{4.37}
\end{equation}
and from (\ref{4.16})
\begin{equation}
n \ge s
\label{4.38}
\end{equation}
But (\ref{4.37}), (\ref{4.38}) contradict (\ref{4.36}). Therefore all 
commutator constraints are satisfied.

\setcounter{equation}{0}
\section{Expectation values and concluding remarks}
We can evaluate the partition function (\ref{1.7}) in the scaling domain
with standard methods (see e.g. \cite{4}, eqs. (181), (182))
\begin{eqnarray}
F(\zeta) &=& \log Z \nonumber \\
&=& {\rm const} + a^{-2\gamma} (1-a^2\zeta)^{-2} \int\limits^{\zeta}
_{a^{-2}} dx(\zeta-x) \nonumber \\
&+& \Bigg\{ \log \bigg[1 + \sum^{\infty}_{m=1} a^{-(m+1)\gamma} 
\begin{array}{c} \\ {\rm Res} \\ z=\infty \end{array} U_m (x;z)\bigg]
\nonumber \\
&+& \log \bigg[1 + \sum^{\infty}_{m=1} a^{-(m+1)\gamma} 
\begin{array}{c} \\ {\rm Res} \\ z=0 \end{array} V_m (x;z)\bigg] \Bigg\}
\label{5.1}
\end{eqnarray}
where $\zeta$ is defined so that (see (\ref{1.25}))
\begin{equation}
\zeta = x(\xi)|_{\xi=1}
\label{5.2}
\end{equation}
In analogy with $D=2$ conformal field theory we may assume maximal holomorphy
for the functions $U_m(x;z)$ and $V_m(x;z)$ in $z$. $V_m(x;z)$ is holomorphic
for $|z| < 1$ with exception of a simple pole at $z=0$, and $U_m(x;z)$
behaves analogously. Moreover, the expansions (\ref{3.1}), (\ref{3.2}) are
assumed to converge in these domains. Then
\begin{equation}
\begin{array}{c} \\ {\rm Res} \\ z=\infty \end{array} U_m (x;z) =
\sum^{\infty}_{r=0} \; \sum_{[n_1,n_2]\in\bbbn^2} U_{mr}^{[n_1,n_2]}(x)
\label{5.3}
\end{equation}
\begin{equation}
\begin{array}{c} \\ {\rm Res} \\ z=0 \end{array} V_m (x;z) =
\sum^{\infty}_{r=0} \;\sum_{[n_1,n_2]\in\bbbn^2} V_{mr}^{[n_1,n_2]}(x)
\label{5.4}
\end{equation}
Due to (\ref{4.14}) for fixed m, the sum over the blocks $[n_1,n_2]$ is
finite. Moreover we can cut off the sum over $r$ at $n_0$ as follows from
the arguments in the preceding section. Then it follows 
\begin{equation}
\begin{array}{c} \\ {\rm Res} \\ z=\infty \end{array} U_1 (x;z) =
\alpha_{1n_0} x + {\rm const}
\label{5.5}
\end{equation}
A second order polynomial in $x$ appears first in
\[ \begin{array}{c} \\ {\rm Res} \\ z=\infty \end{array} U_3 (x;z) \]
The critical exponents $l_1$ and $l_2$ that define the structure of the
quasi-differential operators (see (\ref{1.32}), (\ref{1.33})) are only 
implicitly contained in (\ref{5.3}), (\ref{5.4}), e.g. in the coefficient
$\alpha_{1n_0}$ in (\ref{5.5}).

Finally we want to remark that what we have presented here is not a
mathematical construction, in particular not an existence proof for the
continuous series. This follows from the ``lemma'' mentioned after
(\ref{1.19}). Obviously the commutator (\ref{1.18}) is not diagonal 
automatically. This must be interpreted as an incomplete evaluation of the
Schwinger-Dyson equations. Instead we have imposed the commutator (\ref{1.18})
as a constraint. In the case of the discrete series we proved that the
latter procedure gives equivalent results as a complete evaluation of the
Schwinger-Dyson equations. In the case of the continuous series this is at
most plausible but not yet proved.

\newpage

\setcounter{equation}{0}
\begin{appendix}

{\Large{\noindent \bf{Appendix: Technical considerations}}}

\vspace{0.6cm}
\setcounter{section}{1}

The method applied in sections 2 and 3 uses projection on positive (or 
negative) frequency parts of a Fourier series and a study of asymptotic
expansions in the limit $\varphi \to 0$.

Consider a function $g(z)$ which is holomorphic in the whole complex plane
except a cut from 1 to $\infty$ along the real axis. Then the Taylor
expansion
\begin{eqnarray}
& &g(z) = \sum^{\infty}_{n=0} a_nz^n = \sum^{\infty}_{n=0}a_n r^n 
e^{in\varphi} \\
& &(z = re^{i\varphi}) \nonumber
\label{A.1}
\end{eqnarray}
for $r < 1$ fixed, is a Fourier series with positive frequencies only and
converges exponentially. On the other hand for $r > 1$ fixed
\begin{equation}
g(z) =  \sum^{+\infty}_{n=-\infty} b_n(r)e^{in\varphi}
\label{A.2}
\end{equation}
provided the discontinuity along the cut is integrable
\begin{eqnarray}
& &\int\limits^M_1 d\zeta |g(\zeta+i\epsilon) - g(\zeta-i\epsilon)| < \infty \\
& &(M > r) \nonumber
\label{A.3}
\end{eqnarray}
Then by deformation of the contour
\begin{eqnarray}
a_n &=& \frac{1}{2\pi i} \cdot \oint \frac{dz}{z^{n+1}} g(z) \nonumber \\
&=& \frac{b_n(r)}{r^n} + \frac{1}{2\pi i} \int\limits^r_1 \frac{d\zeta}
{\zeta^{n+1}} [g(\zeta+i\epsilon)-g(\zeta-i\epsilon)]
\label{A.4}
\end{eqnarray}
and
\begin{equation}
a_n = \lim_{r \searrow 1} b_n(r)
\label{A.5}
\end{equation}
In particular this limit vanishes for negative $n$ since
\begin{equation}
a_n = 0 \; {\rm for} \; n < 0.
\label{A.6}
\end{equation}

Consider the typical function
\begin{equation}
g(z) = (1-z)^{\lambda}, \; \lambda \in \bbbc
\label{A.7}
\end{equation}
Then
\begin{equation}
a_n = (-1)^n {\lambda \choose n}
\label{A.8}
\end{equation}
\begin{equation}
b_n(r) = \frac{\sin \pi \lambda}{\pi} \frac{r^{\lambda}}{\lambda-n} \,_2F_1
(-\lambda, -\lambda+n; - \lambda+n+1; \frac1r)
\label{A.9}
\end{equation}
Provided
\begin{equation}
Re \lambda + 1 > 0
\label{A.10}
\end{equation}
we can verify (\ref{A.5}).

Now consider $\lambda = -1$. Then
\begin{equation}
a_n = \left\{ \begin{array}{lr}
 1&n \ge 0 \\ 0 & n < 0 \end{array} \right.
\label{A.11}
\end{equation}
\begin{equation}
\lim_{r\searrow1} b_n(r) = \left\{ \begin{array}{lr}
- 1 & n < 0 \\ 0 & n \ge 0 \end{array} \right.
\label{A.12}
\end{equation}
so that
\begin{equation}
\sum^{+\infty}_{n=-\infty} (a_n-b_n(1+))e^{in\varphi} = 2\pi\delta(\varphi)
\label{A.13}
\end{equation}
In the asymptotic expansion at $\varphi \to 0$ the r.h.s. of (\ref{A.13}) is
unobservable. A similar situation arises if $\lambda$ is equal any negative 
integer.

If
\begin{equation}
Re \lambda + 1 \le 0, \; - \lambda \notin \bbbn
\label{A.14}
\end{equation}
then we can perform the limit $r \searrow 1$ on (\ref{A.9}) after subtraction
of a diverging expression which renders the result analytic in $\lambda$
(``analytic regularization''). This diverging subtraction term consists of
local distributions 
\[ \frac{d^s}{d\varphi^s} \delta(\varphi) \]
multiplied with diverging coefficients
\[ \sim \left(1-\frac1r \right)^{\lambda+1-s} \]
In any case instead of (\ref{A.5}) we have
\begin{equation}
a_n = \lim_{r \searrow 0} \; \{b_n(r) - \mbox{ asymptotically unobservable 
terms} \} 
\label{A.15}
\end{equation}

As an example we prove (\ref{3.8}). Derivation of (\ref{3.3}) gives
\begin{equation}
\left( - \frac1z (1-z)^{l_1} - l_1(1-z)^{l_1-1)}\right)_{\uparrow,+}
\label{A.16}
\end{equation}
on the l.h.s. and
\begin{equation}
\left(\left( 1 + \frac{l_2}{z(1-\frac1z)} \right) \sum^{\infty}_{k=1} k
f^c_k z^k (1-\frac1z)^{kl_2} \right)_{\downarrow,+}
\label{A.17}
\end{equation}
on the r.h.s. The factor in front of the sum of (\ref{A.17}) produces only
nonpositive frequencies and can therefore be written as
\begin{eqnarray}
\left(\left( 1 + \frac{l_2}{z(1-\frac1z)} \right)_{\downarrow}
\left( \sum^{\infty}_{k=1} k f^c_k z^k (1-\frac1z)^{kl_2} \right)
_{\downarrow,+} \right)_+ \nonumber \\
= \left(\left( 1 + \frac{l_2}{z(1-\frac1z)} \right)_{\downarrow}
\left( \sum^{\infty}_{k=1} c_n (1-z)^{l_1+n}\right)_{\uparrow} \right)_+
\nonumber \\
+ \mbox{ irrelevant terms} \qquad \qquad 
\label{A.18}
\end{eqnarray}
Now (\ref{A.13}) implies 
\begin{equation}
\left( \frac{1}{z(1-\frac1z)} \right)_{\downarrow} = -
\left( \frac{1}{1-z} \right)_{\uparrow} + 2 \pi \delta(\varphi)
\label{A.19}
\end{equation}
so that we can continue (\ref{A.18}) to
\begin{eqnarray}
\left( \sum^{\infty}_{n=0} c_n \left[(1-z)^{l_1+n} - l_2(1-z)^{l_1+n-1}
\right]\right)_{\uparrow,+} 
\nonumber \\
+ \mbox{ irrelevant terms} \qquad \qquad
\label{A.20}
\end{eqnarray}
where the latter include a holomorphic  part at $z = 1$ with an infinite
constant (if $Re \,l_1 < 0$) and eventually a first order pole at $z=1$.
Equating (\ref{A.16}) and (\ref{A.18}) we obtain (\ref{3.8}).
\end{appendix}

\section*{Acknowledgement}
S.B. would like to thank the German Academic Exchange Service (DAAD) for 
financial support.


\newpage
\begin{thebibliography}{99}

\bibitem{1} M.L. Metha, Random matrices and the statistical theory of energy levels,
Acad. Press, New York 1967.
\bibitem{2} C. Itzykson, J.-M. Drouffe, Statistical field theory, 2 Vols, Camb. Univ.
Press, Cambridge 1989, Section 10.3
\bibitem{3}E. Br'zin, V.A. Kazakov, Phys. Lett. B236 (1990) 144; \\
D.J. Gross, A.A. Migdal, Phys. Rev. Lett. 64 (1990) 127; \\
M.R. Douglas, S.H. Shenker, Nucl. Phys. B335 (1990) 635; \\
Proceedings of 1990 CargŠse workshop on ``Random surfaces and quantum gravity'',
Eds. O. Alvarez, E. Marinari, P. Windey. NATO ASI Series B: Physics Vol. 262
(1992); \\
P. di Francesco, P. Ginsparg, J. Zinn-Justin, Physics Reports \underline{254}
 (1995) 1-133.
\bibitem{4} S. Balaska, J. Maeder, W. R\"uhl, Nucl. Phys. B 486 (1997) 673.
\bibitem{5} M.R. Douglas, Phys. Lett B238 (1990) 176; \\
P. Ginsparg, M. Goulian, M.R. Plesser, J. Zinn-Justin, Nucl. Phys.
B342 (1990) 539.
\bibitem{6} I.M. Gelfand, L.A. Dikii, Funks. Anal. Prilozhen., \underline{10},
4 (1976) 13.
\bibitem{7} D. Bessis, Commun. Math. Phys. 69 (1979) 147; \\
D. Bessis, C. Itzykson, J.-B. Zuber, Adv. Appl. Math. 1
(1980) 109.
\bibitem{8} M.L. Metha, Commun. Math. Phys. 79 (1981) 327; \\
S. Chadha, G. Mahoux, M.L. Mehta, J. Phys. A 14 (1981) 579. 
\end{thebibliography}
\end{document}